\documentclass{article}

\PassOptionsToPackage{numbers, compress}{natbib}
\usepackage[final]{neurips_2023_ml4ps}

\usepackage[utf8]{inputenc} % allow utf-8 input
\usepackage[T1]{fontenc}    % use 8-bit T1 fonts
\usepackage{hyperref}       % hyperlinks
\usepackage{url}            % simple URL typesetting
\usepackage{booktabs}       % professional-quality tables
\usepackage{amsfonts}       % blackboard math symbols
\usepackage{nicefrac}       % compact symbols for 1/2, etc.
\usepackage{microtype}      % microtypography
\usepackage{xcolor}         % colors
\usepackage[pdftex]{graphicx}

\newcommand{\xHIv}{{\left< x_{\rm HI}\right>_{\rm v}}}
\newcommand{\Lya}{{\rm Ly}\alpha}
\newcommand{\CHIMP}{~h^{-1}{\rm~cMpc}}

\definecolor{red}{rgb}{1,0,0}

\title{Learning Reionization History from Quasars with Simulation-Based Inference}

\author{
  Huanqing Chen\\
  Canadian Institute for Theoretical Astrophysics\\
  University of Toronto\\
  Toronto, ON M5S 3H8, Canada\\
  \texttt{hqchen@cita.utoronto.ca} \\
  \And
  Joshua Speagle \\
  Department of Statistical Sciences \\
  Department of Astronomy \& Astrophysics \\
  University of Toronto \\
  Toronto, ON M5S 3H4, Canada \\
  \texttt{j.speagle@utoronto.ca} \\
  \And
  Keir K. Rogers \\
  Dunlap Institute for Astronomy \& Astrophysics\\
  University of Toronto \\
  Toronto, ON M5S 3H4, Canada \\
  \texttt{keir.rogers@utoronto.ca} \\
}

\begin{document}

\maketitle

\begin{abstract}
Understanding the entire history of the ionization state of the intergalactic medium (IGM) is at the frontier of astrophysics and cosmology. A promising method to achieve this is by extracting the damping wing signal from the neutral IGM. As hundreds of redshift $z>6$ quasars are observed, we anticipate determining the detailed time evolution of the ionization fraction with unprecedented fidelity. However, traditional approaches to parameter inference are not sufficiently accurate.

We assess the performance of a simulation-based inference (SBI) method to infer the neutral fraction of the universe from quasar spectra. The SBI method adeptly  exploits the shape information of the damping wing, enabling precise estimations of the neutral fraction $\xHIv$ and the wing position $w_p$.
Importantly, the SBI framework successfully breaks the degeneracy between these two parameters, offering unbiased estimates of both.
This makes the SBI superior to the traditional method using a pseudo-likelihood function. 
We anticipate that SBI will be essential to determine robustly the ionization history of the Universe through joint inference from the hundreds of high-\(z\) spectra we will observe.
\end{abstract}

\section{Introduction}
\begin{figure}
  \includegraphics[width=0.99\linewidth]{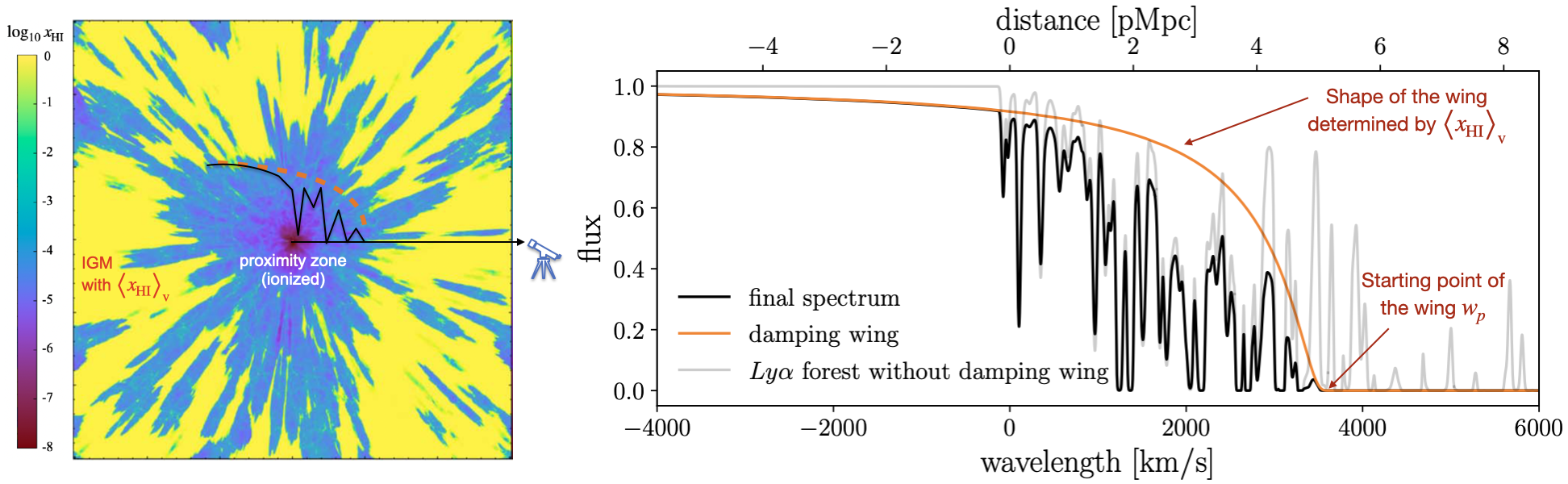}
  \vspace{-0.2cm}
  \caption{Left: A slice of a quasar proximity zone simulation \cite{chen2020}. Blue represents ionized gas while yellow indicates neutral gas. It also contains a cartoon illustrating the quasar spectrum: the blue region towards the observer corresponds to a $\Lya$ forest, enveloped by a damping wing (orange) arising from the neutral gas. Right: A spectrum simulated with the method described in Sec. \ref{sec:simulator}. The final spectrum (black) is a multiplication of the $\Lya$ forest (grey) and the damping wing (orange).\label{fig:illustration}   }
\end{figure}

Understanding the evolution of the neutral fraction of the universe is a central topic in the study of cosmic reionization. 
Up to now, the most successful and well-known constraint has been from analyzing the Thompson scattering signal in the Cosmic Microwave Background (CMB)\cite{planck2016,planck2020,act2011,spt2012}. However, due to its integrated effect in the CMB, constraining the detailed time evolution of the cosmic neutral fraction is challenging. Another approach is to analyze the effect of the neutral intergalactic medium (IGM) on spectra from quasars directly living in the epoch of reionization (Fig. \ref{fig:illustration}).
With hundreds of high-redshift quasars discovered and followed-up spectroscopically \cite{dodorico2023,euclid2019,tee2023}, this method could potentially surpass the CMB method in the near future, if the information can be extracted reliably.

Neutral gas in the IGM forms a damping wing on quasar spectra. Collectively, many neutral patches in a quasar sightline create a characteristic damping wing for a given $\xHIv$ with a small scatter arising from cosmic variance \cite{chen2023,keating2023}.
If we could unambiguously measure the entire $\Lya$ damping wing, constraining the neutral fraction would be straightforward. However, the residual neutral hydrogen inside the quasar proximity zone (i.e. the region where the gas is ionized by the quasar) could cause significant absorption, rendering the damping wing profile hard to recover.

Nevertheless, such structure (i.e. the $\Lya$ forest) inside the proximity zone is dictated by the cosmic large-scale structure, which is straight forward to model \cite{rauch1998}. In Fig. \ref{fig:illustration}, we illustrate the two main components of a proximity zone spectrum. The left panel shows a snapshot of a radiative transfer cosmological simulation with a quasar in the center \cite{chen2020}. The quasar generates a large number of photons ionizing the gas (blue region) in its surroundings. The gas outside the quasar ionization front is not impacted, thus remaining neutral (yellow region). If the universe is already ionized, there will be no neutral patches. In this scenario, the quasar spectrum has only one component: the $\Lya$ forest in the proximity zone (grey line in the right panel of Fig. \ref{fig:illustration}). However, when reionization is not complete, there will be a damping wing component (orange envelope) that arises outside the quasar proximity zone and suppresses all the flux at smaller wavelengths. The position of the damping wing is determined by the quasar's past activities, but the shape of the damping wing is determined by the global neutral fraction \cite{chen2023,keating2023}.  Measuring this envelope accurately is therefore key to measuring the neutral fraction.

Due to the complexity of the $\Lya$ forest, it is challenging to explicitly write down the likelihood function of the flux at each wavelength (i.e. pixel). Previous studies \cite{davies2018,yang2020} thus used a ``pseudo-likelihood'' function to compress all of this information into one number, leading to degraded and/or biased parameter constraints. Simulation-based inference (SBI) has emerged as a rapidly advancing technique specifically aiming to solve inference problems with intractable likelihoods. We thus set out to assess whether SBI using conditional normalizing flows can improve the inference compared to the traditional pseudo-likelihood method.

\section{Data} \label{sec:simulator}

To accurately model the details in the proximity zone, we need hydrodynamical cosmological simulations and radiative transfer. However, simulating spectra with radiative transfer on-the-fly can be expensive. To reduce the simulating time, we develop a simulator that makes composite spectra using two pre-calculated datasets: proximity zone $\Lya$ forests and damping wings.

We create a set of proximity zone $\Lya$ forests by post-processing the Cosmic Reionization On Computers (CROC) simulations \cite{gnedin2014}. In each of the six $40 \CHIMP $ simulation boxes, we first select the 20 most massive halos. We then draw 50 sightlines with random orientations centered on each halo to sample the cosmological environments around quasars, and use 1D radiative transfer modelling \cite{chen2021a} to create a large set of $\Lya$ spectra ranging from -4000 km/s to 8000 km/s.

To model damping wing components, we use the simulated damping wing database created by \cite{chen2023}. This database samples 10,000 sightlines traveling through different neutral patches in a universe with a given $\xHIv$, capturing realistic scatter in the damping wing profile. The database is evenly spaced across 21 global neutral fractions ranging from $\xHIv=0.995$ to $\xHIv=0.008$.
For each $\xHIv$ and the starting point of the damping wing $w_p$, we then randomly draw one $\Lya$ forest and one damping wing with the closest $\xHIv$. We then move the wing to $w_p$ and multiply the flux with the transmitted proportion of the wing on the red side and with zero on the blue side. We show an example of a simulated composite spectrum in the left panel of Fig. \ref{fig:example}.  

Real spectra usually have noise of varying levels. As a result, previous studies often have binned the data as a function of wavelength to reduce the effective noise per binned pixel \cite{davies2018}. In order to make realistic comparison between SBI and traditional pseudo-likelihood approach described in \cite{davies2018}, we also bin our simulated quasar spectra to a similar resolution (500 km/s) over the entire spectral range in our subsequent tests (see the blue line in the left panel of Fig. \ref{fig:example} as an example). 
% We anticipate the performance of SBI to improve if this binning scheme is subsequently relaxed to allow for higher wavelength resolution. 
{In the Supplementary Material, we show how the result changes with different noise levels and number of spectra used.}

% both the shape of damping wing component and the large-scale structure to generate proximity zone Lya forest. 

% The spectra that are used to infer the neutral fraction is the region around the Lya line. Such a spectrum has of two important components: the damping wing arised from neutral gas outside the quasar impact region (proximity zone) and the absorption inside the proximity zone by residual neutral gas in the cosmic large-scale structure.

% The questions we want to answer are:
% 1. If we use the best quality spectra currently available, what inference we would get? Is it biased?
% 2. What about we choose another resolution (500km/s) binned?
% 3. If we use multiple sightlines, can we improve the results?

\section{Methods} \label{sec:sbi}

\subsection{Simulation-Based Inference}
% \subsubsection{Architecture:}
We use the default Neural Posterior Estimation (NPE) \citep{papamakarios2016, greenberg2019} framework as implemented by the \texttt{sbi}\footnote{https://www.mackelab.org/sbi/. {We use the ``SNPE\_C'' class with a single round.} }package without any additional embedding network. {The NPE aims to find the neural network (a density estimator) $\phi$ which maximizes
    $\Sigma_{i=0}^{i=N} \ln P(\theta_i|x_i, \phi)$, where $(\theta_i, x_i)$ are independent (parameter, simulation) pairs generated from the prior.}
For the density estimator, we adopt Masked Autoregressive Flow (MAF) \cite{papamakarios2017} trained on 100,000 simulated spectra. We run a small grid search for two hyper-parameters: the number of hidden features $n_{\rm hf}$ = ($5, 10, 20$) and number of transformations $n_{\rm t}=(5, 10, 20)$. We found that all the combinations return sensible results with clearly decoupled parameter constraints. We therefore opt to use MAF with the simplest hyper-parameter combination $(n_{\rm hf}, n_{\rm t}) = (5,5)$, which converged after 277 epochs.\footnote{All computation was done on the SciNet NIAGARA supercomputer using one CPU node with 40 cores. The total training of our final model took 73 minutes of CPU time.} %\footnote{https://www.scinethpc.ca/niagara/}

\subsection{Pseudo-likelihood Approach}
In \cite{davies2018}, a pseudo-likelihood $\tilde{L}(x|\theta)$ function is proposed to enable a Bayesian inference of $\xHIv$ from quasar spectra. The pseudo-likelihood function is calculated by simply multiplying the likelihood for each wavelength bin assuming they are all independent:
\begin{equation}\label{eq}
    \tilde{L}(x|\theta) = \prod_{i=1}^{n_{\rm bins}} \tilde{L}_i(x_i|\theta) 
\end{equation}
Similar to \cite{davies2018}, we create a uniform grid of parameters with 21 bins in $\xHIv$ and 16 bins in $w_p$ from 0 to 8000 km/s. We use the simulator described in Sec. \ref{sec:simulator} to generate 1000 spectra for each parameter pair ($\xHIv$, $w_p$). Then we use the distribution of flux values to estimate the flux probability distribution function (PDF) at each bin in order to evaluate the likelihood $\tilde{L}_i(x_i|\theta)$.\footnote{The flux PDF is estimated using a histogram with 50 bins, although we find slightly changing the number of bins {or using kernel density estimation with Gaussian kernels} does not have significant impact on the final results.}

\section{Results}

\begin{figure}
  \includegraphics[width=0.32\linewidth]{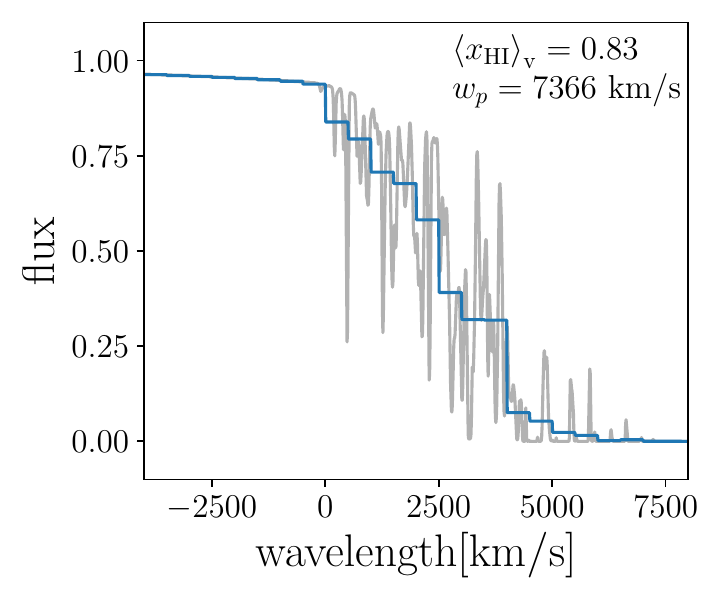}
  \hspace{-0.29cm}
\includegraphics[width=0.345\linewidth]{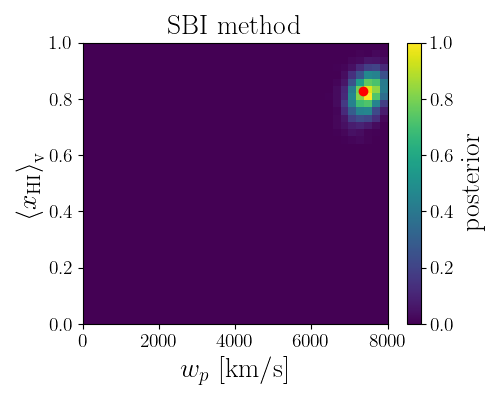}
  \hspace{-0.29cm}
\includegraphics[width=0.345\linewidth]{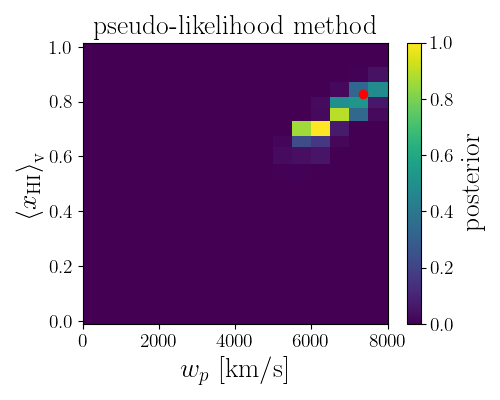}
  \vspace{-0.2cm}
  \caption{Left: An example of simulated spectra. The grey curve represents the high-resolution spectrum and the blue curve is the binned spectrum (bin size = 500 km/s) used as the summary statistics. Middle: Unnormalized posterior of the parameters infered by the SBI. Right: Unnormalized posterior using the traditional method with the ``pseudo-likelihood" function. The right panel is at a lower resolution because the pseudo-likelihood is calculated on a uniform grid. The red point indicates the true parameters.\label{fig:example} \label{fig:bias}
  }
\end{figure}

{We use a uniform prior for both $\xHIv$ and $w_p$.} In the middle and right panels of Fig. \ref{fig:example}, we show SBI and the pseudo-likelihood distributions for the example spectrum on the left panel. Compared with the SBI result in the middle panel, the pseudo-likelihood method shows stronger, more extended degeneracies between the two parameters and lower resolution due to the coarser input parameter grid, reliance on a single summary statistic $\tilde{L}$, and ignorance of potential intrapixel covariances. 
This degeneracy in the pseudo-likelihood method can be especially severe when $w_{p, \rm true}$ is large: for $w_p> w_{p, \rm true}$, many bins in Eq.\ref{eq} have non-zero $\tilde{L}_i(x_i|\theta)$, which can be compensated with a stronger damping wing (smaller $\xHIv$). 
On the other hand, SBI could extract the shape information of the damping wing from correlated pixels to break the degeneracy between the two parameters.

% We use an architecture without any embedding net. For the density estimator, we first explore MAF. 
We also performed a probability calibration check \cite{talts2018} using the ranks of the estimated quantiles for 1000 test spectra to check whether there are biases across both sets of constraints (Fig. \ref{fig:bias}). We find that the rank distribution is consistent with a uniform distribution for both parameters $\xHIv$ and $w_p$: comparing the rank distribution with the uniform distribution using the Kolmogorov–Smirnov (K-S) test, we obtain a $p$-value of 0.16 and 0.56 for $\xHIv$ and $w_p$, respectively. This implies that there is no strong reason to believe that the collection of SBI PDFs do not accurately capture the scatter in true values given our estimated uncertainties.

% This results indicate that using SBI can indeed significantly improve the constraints. 
From left to right, the blue histograms in Fig. \ref{fig:bias} show the rank distribution, bias distribution and (two-sided) 68\% interval of the $\xHIv$ constraint from the SBI method, respectively. For a single spectrum, we find the root mean square (RMS) of bias in $\xHIv$ to be $0.06$, which is smaller than the RMS of 68\% scatter (0.12).
% This shows that this SBI framework produces unbiased results from another perspective. 
Moreover, the typical scatter is approximately equal to the typical cosmic variance of the damping wing \cite{chen2023}. This demonstrates that SBI not only produces an unbiased inference of $\xHIv$, but that these constraints are close to optimal.

\begin{figure}
\includegraphics[width=0.33\linewidth]
{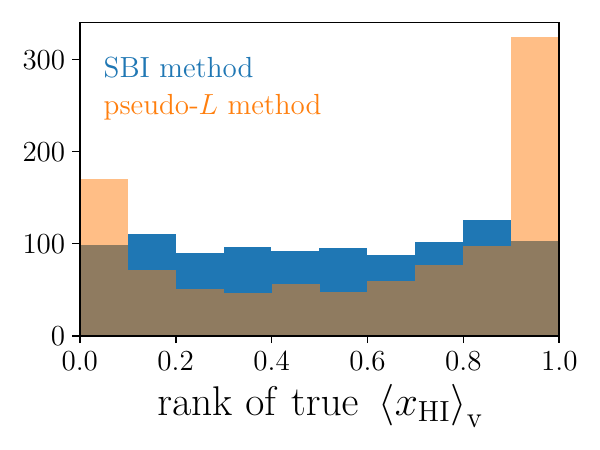}
\includegraphics[width=0.33\linewidth]{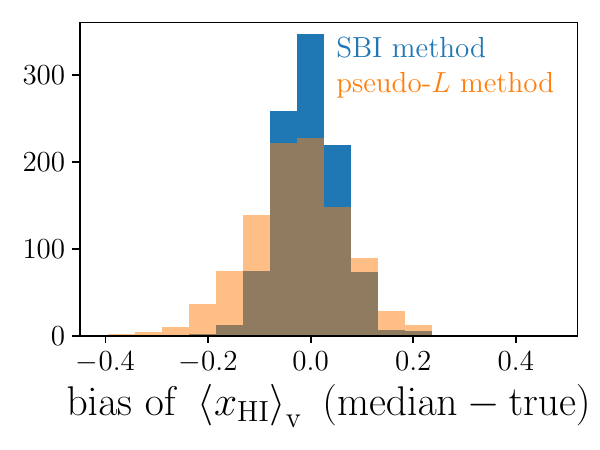}
\includegraphics[width=0.33\linewidth]{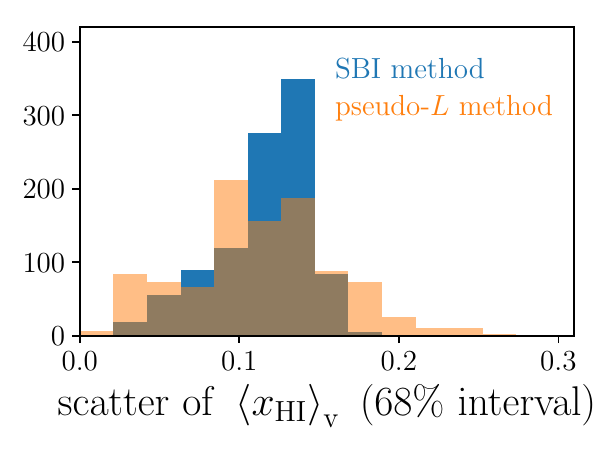}
  \caption{Rank distribution (left), bias (middle) and scatter (right) of the SBI method (blue) and the traditional method (orange). The number of test data points for both methods are 1000.\label{fig:bias}
  }
\end{figure}

% \subsection{Comparison with the traditional method}
% In \cite{davies2018}, a pseudo-likelihood $\tilde{L}(x|\theta)$ function is proposed to enable a Bayesian inference of $\xHIv$ from quasar spectra. The pseudo-likelihood function is calculated as the multiplication of the likelihood of each bin:
% \begin{equation}\label{eq}
%     \tilde{L}(x|\theta)= \prod_{i=1}^{n_{\rm bins}} \tilde{L}(x_i|\theta) 
% \end{equation}

% This implicitly assumes that each pixel does not correlate with each other, which is not strictly correct. Here we assess this traditional method and compare it with the SBI method.

% The degeneracy is not as severe as one would expect though, because for $w_p > w_{p, \rm true}$, Eq. \ref{eq} results in multiplication of many extremely small values for bins whose wavelength $>w_{p, \rm true}$ , effectually heavily penalize the likelihood where $w_p> w_{p, \rm true}$.

On the other hand, the traditional pseudo-likelihood method introduces significant bias. We show the rank distribution, bias and scatter of the inference on $\xHIv$ in the faint orange histogram in Fig. \ref{fig:bias}. The ``U'' shape clearly demonstrates that the traditional method under-predicts the uncertainties in the inference. The RMS of the bias is $0.11$, larger than the SBI method. The RMS of the scatter of the pseudo-likelihood method (0.124) is similar to SBI (0.120), but the distribution is noticeably skewed towards both smaller and large values compared to SBI.

% \begin{figure}
% \includegraphics[width=0.48\linewidth]{42.png}
%   \includegraphics[width=0.48\linewidth]{bin500kmps_5spec_rank_cdf_maf_5_5_2.pdf}
%   \caption{Left: constraints for xHIv and five wps when doing SBI for five quasars at the same redshift. xHIv constraint is more precise. Right: rank CDF.}
% \end{figure}

\section{Discussion}

We demonstrated that using SBI, we can more accurately constrain the neutral fraction of the universe using individual quasar spectra. SBI performs systematically better than traditional pseudo-likelihood methods, returning PDFs that are more accurate, and with more realistic uncertainties over our parameters of interest.

For an individual spectrum, the traditional method shows larger bias than the SBI method. Such bias could compromise our results when inferring from multiple spectra. Up to now, there are hundreds of quasars detected at $z>6$ with high-quality spectra, and one key step forward is to build an unbiased methodology to jointly constrain the global neutral fraction $\xHIv$. From this perspective, the SBI method is clearly more desirable. The SBI can also be straight-forwardly implemented in such a multiple-spectra case by  simply concatenating multiple spectra as input data\footnote{See Supplementary Material.}.
% We have tested a 5-spectra inference case and found that the RMS of the bias is reduced to 0.03 while the scatter is reduced to 0.06\footnote{An example and the bias studies can be found in the supplementary material.}. 

\section{Broader Impact}

This is a pioneering study on using SBI with state-of-the-art radiative transfer cosmological simulations. The results show the promising potential of using quasar spectra to measure the detailed time evolution of the IGM ionization state, which is crucial for both cosmology and astrophysics. In the future, we will incorporate various realistic factors, especially continuum uncertainties \cite{greig2017,davies2018predicting,vdurovvcikova2020,liu2021,sun2023} into the SBI framework and develop end-to-end machinery to uncover the reionization history.
The methodology presented in this study can also be extended to detect and characterize different absorption systems such as Lyman Limit Systems or Damped $\Lya$ Absorbers in the $\Lya$ forest in quasar spectra \citep{Garnett:2016awq,Rogers:2017bmq,Rogers:2017eji}, which is crucial for understanding structure formation.

\begin{ack}
This study has been supported by the Natural Sciences and Engineering Research Council of Canada (NSERC), funding references \#DIS-2022-568580 and \#RGPIN-2023-04849.
The Dunlap Institute is funded through an endowment established by the David Dunlap family and the University of Toronto.
% Do {\bf not} include this section in the anonymized submission, only in the final paper. You can use the \texttt{ack} environment provided in the style file to autmoatically hide this section in the anonymized submission.
\end{ack}

% Authors may wish to optionally include extra information (complete proofs, additional experiments and plots) in the appendix. All such materials should be part of the supplemental material (submitted separately) and should NOT be included in the main submission.

% \begin{figure}\label{fig:5spec_example}
%   \includegraphics[width=0.99\linewidth]{figs/3_corner_5spec.png}
%   \caption{An example of SBI result of simultaneously constraining $\xHIv$ and $w_p$ using 5 quasar spectra at the same redshift.}
% \end{figure}

% \section*{References}

% References follow the acknowledgments in the camera-ready paper. Use unnumbered first-level heading for
% the references. Any choice of citation style is acceptable as long as you are
% consistent. It is permissible to reduce the font size to \verb+small+ (9 point)
% when listing the references.
% Note that the Reference section does not count towards the page limit.
\medskip
{\small
\bibliographystyle{plain}
\bibliography{neurips_2023}
}

\clearpage

\section{Supplementary Material}

% \vspace{10cm}

% \begin{figure}[h!]\label{fig:5spec_example}
  % \includegraphics[width=1\linewidth]{figs/5spec_example.png}
  % \caption{An example of the SBI results when simultaneously constraining  $\xHIv$ and $w_p$ using 5 quasar spectra at the same redshift. Upper panels: The first five panels show the five test spectra. The rightmost panel shows the parameter constraint for $\xHIv$. Lower panels: each panel shows the 2D histogram of the posterior distribution of $\xHIv$ and the corresponding $w_p$ of the spectrum plotted above.}
% \end{figure}

% \begin{figure}[h!]\label{fig:bias_5spec}
% \includegraphics[width=0.33\linewidth]{figs/SBI_rank_5spec.pdf}
%   \includegraphics[width=0.33\linewidth]{figs/SBI_bias_5spec.pdf}
%   \includegraphics[width=0.33\linewidth]{figs/SBI_scatter_5spec.pdf}
%   \caption{Rank distribution (left), bias (middle) and scatter (right) of the SBI on $\xHIv$ using 5 quasar spectra at the same redshift. The number of test data points is 1000. The $p-$value of the Kolmogorov–Smirnov test between the rank distribution and the uniform distribution is $0.08$.}
% \end{figure}

\subsection{Spectra with Noise}
\begin{figure*}[!h]
    \includegraphics[width=0.33\textwidth]{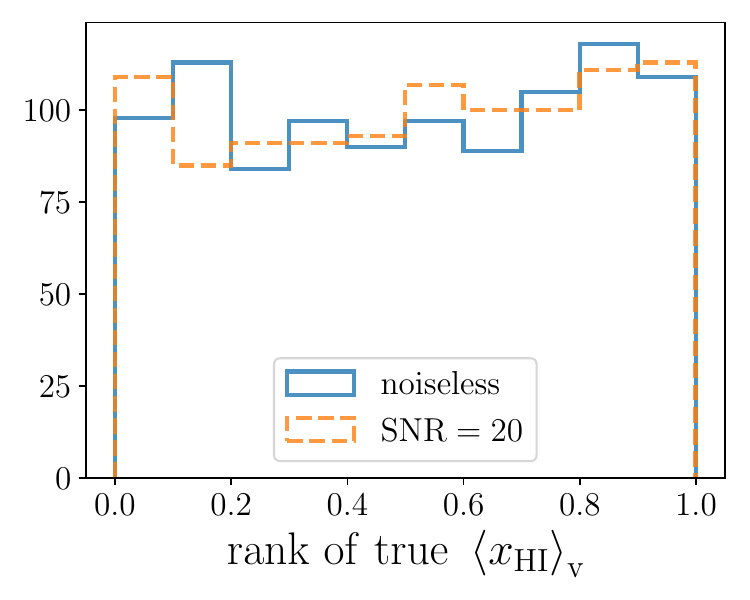}
    \includegraphics[width=0.33\textwidth]{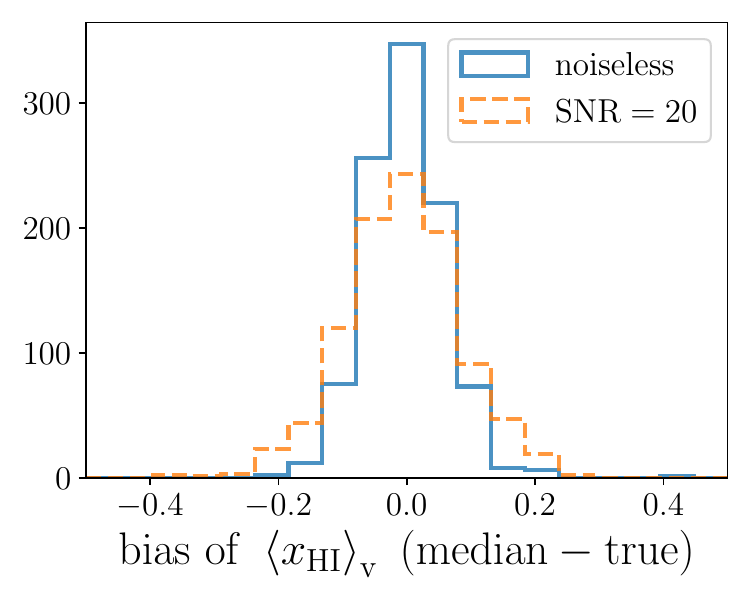}
    \includegraphics[width=0.33\textwidth]{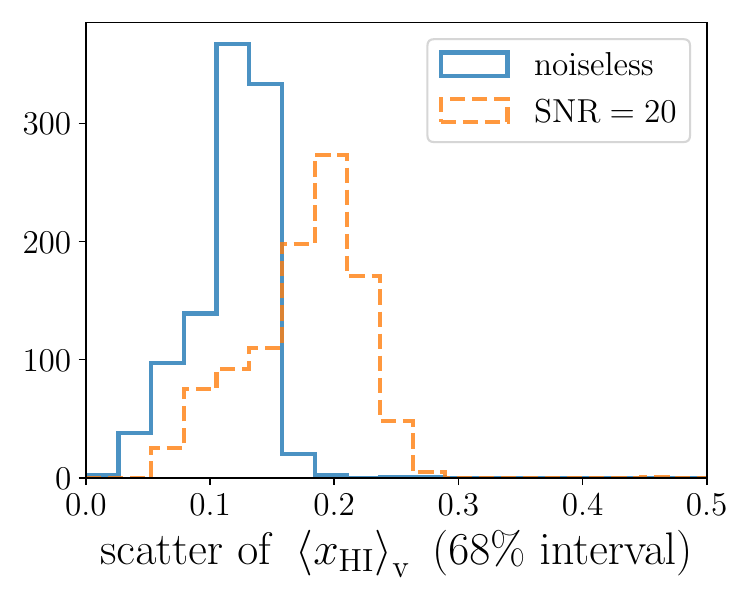}
    \includegraphics[width=0.33\textwidth]{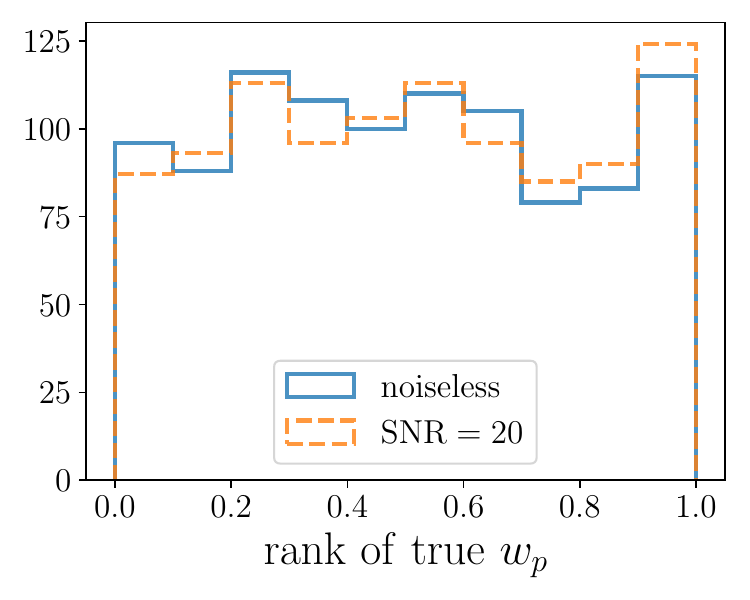}
    \includegraphics[width=0.33\textwidth]{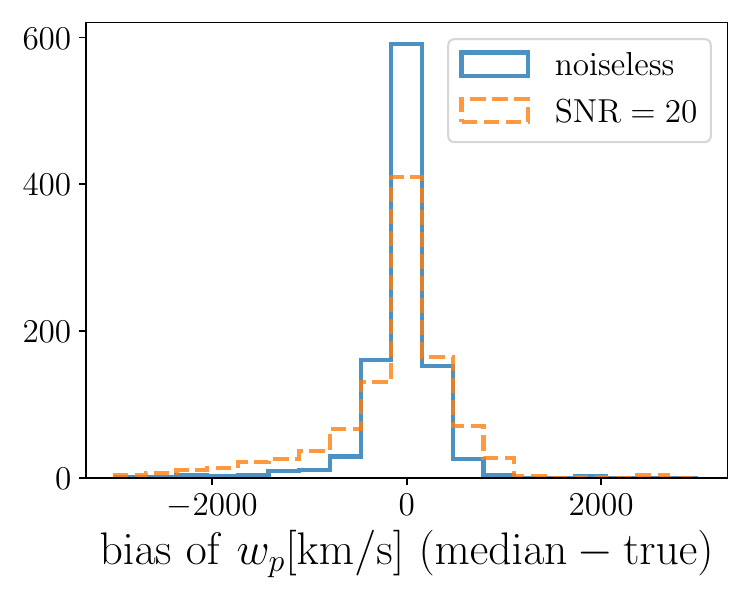}
    \includegraphics[width=0.33\textwidth]{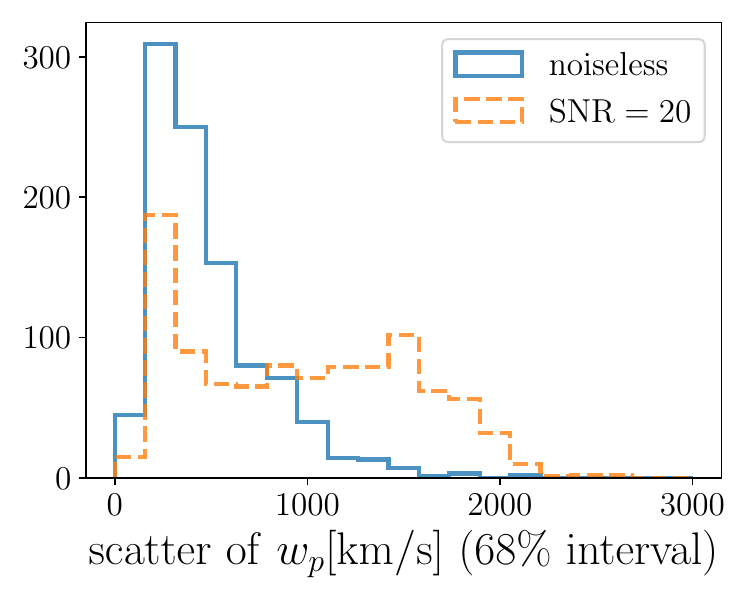}
    \caption{Comparison of rank distribution (left), bias (middle) and scatter (right) of the noiseless case (blue solid) and the SNR=20 case (orange dashed), both using SBI method. The number of test data points for both methods are 1000. \label{fig:noise}}
\end{figure*}

Real spectra always have some level of noise. Currently, the sample with the highest quality has an average signal-to-noise (SNR) $\sim 20$ per $10$ km/s \citep{dodorico2023}. We thus test how adding such noise impacts the results. In each spectrum, we first add an uncorrelated Gaussian noise with zero mean and a standard deviation of $5\%$ of the continuum level, and then bin it to $500$ km/s per bin.
We again train a MAF with 5 hidden features and 5 transformations and test the results on a separate set of 1000 spectra with the same noise level. The simulation-based calibration shows the rank distributions are consistent with a uniform distribution for both $\xHIv$ (K-S test $p$-value $=0.13$) and $w_p$ ($p$-value $=0.41$), indicating 
the uncertainties inferred from the SBI
accurately capture the scatter in true values.
The bias of each individual spectrum is increased slightly compared with the noiseless case: the RMS increases from $0.06$ to $0.09$ for $\xHIv$, and from $505$ km/s to $775$ km/s for $w_p$, while the RMS scatter is increased from $0.12$ km/s to $0.18$ km/s for $\xHIv$, and from $720$ km/s to $1120$ km/s for $w_p$ (see also Fig. \ref{fig:noise}).

\subsection{Scaling with Number of Spectra}
Using multiple spectra at the same redshift could increase the constraint on $\xHIv$. This is especially promising as we will have more reionization epoch quasars observed in the near future \cite{euclid2019, tee2023}. Here we test how the inference scales with the number of spectra with different noise levels. Because we are primarily interested in $\xHIv$, here we only train it to infer the single parameter $\xHIv$. The input data for multiple spectra inference is constructed by simply concatenating the spectra together into a single 1D array.

\begin{figure*}
    \includegraphics[width=0.33\textwidth]{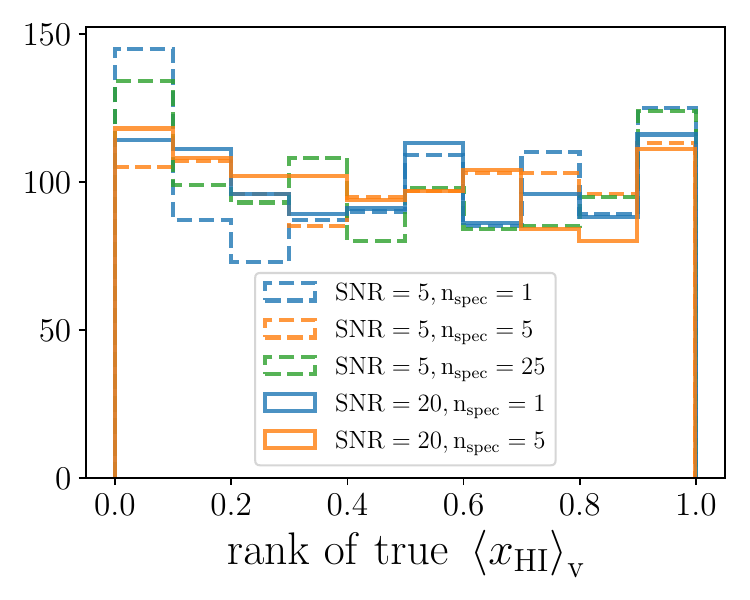}
    \includegraphics[width=0.33\textwidth]{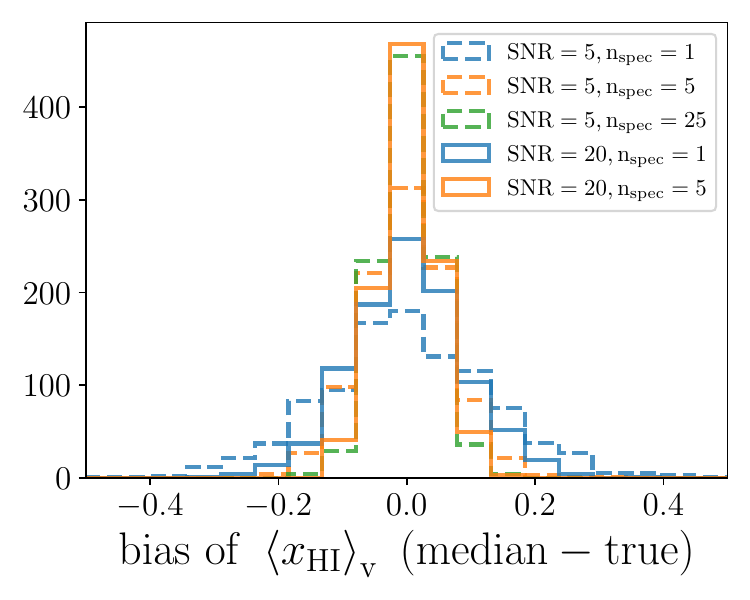}
    \includegraphics[width=0.33\textwidth]{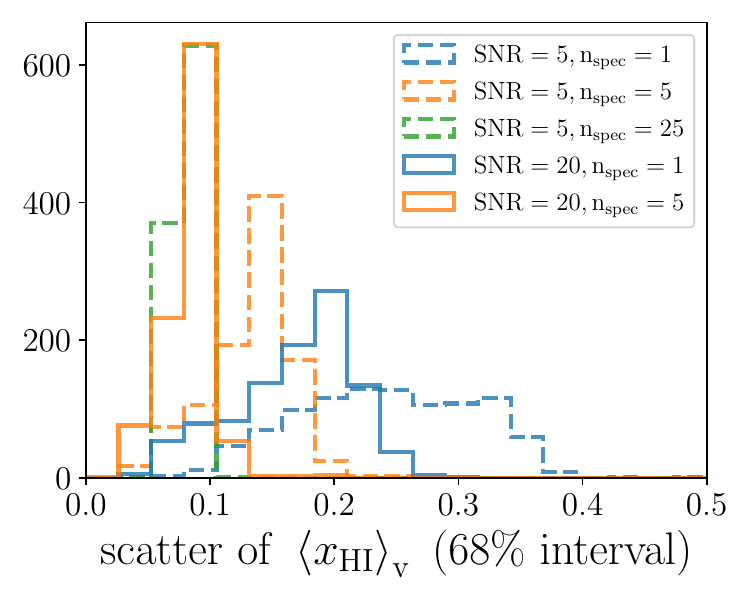}
    \caption{Rank distribution (left), bias (middle) and scatter (right) of cases of different number of spectra with different SNRs.\label{fig:nspec}}
\end{figure*}

We test a few realistic cases: SNR=5 for 1, 5 and 25 spectra and SNR=20 for 1 and 5 spectra. We use the same hyper-parameters ($n_{\rm hf}=5$, $n_{\rm t}=5$) and find that except for the single spectrum with SNR=5 case, all the others return an unbiased inference based on the K-S test on rank distribution with uniform distribution: the single SNR=5 spectrum case has $p$-value=0.008 while others all have $p$-value>0.05. For the single SNR=5 spectrum case, we further test other hyper-parameter combinations with $n_{\rm hf}$ varying between (3,5,10,20) and $n_{\rm t}$ varying between (2,3,5,10,20), but the highest $p$-value returned is still only $0.03$ (from combination $n_{\rm hf}=3$, $n_{\rm t}=2$). This indicates that using a single spectrum with only SNR=5, it is very challenging to obtain an accurate $\xHIv$ constraint. In Fig. \ref{fig:nspec} we show the test results for each case, all trained with $n_{\rm hf}=5, n_{\rm t}=5$. 
For SNR=5 spectra, the RMS bias improves from 0.13, 0.069, to 0.044, and the RMS scatter improves from 0.25, 0.14 and 0.081 as we increase the number of spectra from 1, 5, to 25, respectively.
On the other hand, for SNR=20 spectra, the RMS bias improves from 0.089 to 0.045 while the RMS scatter from 0.18 to 0.088 as we increase the number of spectra from 1 to 5.

\end{document}